\documentclass[iop]{emulateapj}   

\slugcomment{draft version \today, accepted for publication in the Astrophysical Journal}

\usepackage{comment}
\usepackage[hidelinks]{hyperref}
\usepackage{amsmath}

\shorttitle{}
\shortauthors{}

\begin{document}

\title{ALMA Reveals Weak [N{\sc ii}] Emission in ``Typical" Galaxies and Intense Starbursts at $z=5$--6}

\author{Riccardo Pavesi\altaffilmark{1}$^\dagger$, Dominik A. Riechers\altaffilmark{1}, Peter L. Capak\altaffilmark{2}, Christopher L. Carilli\altaffilmark{3,4}, Chelsea E. Sharon\altaffilmark{1}, Gordon J. Stacey\altaffilmark{1}, Alexander Karim\altaffilmark{5}, Nicholas Z. Scoville\altaffilmark{6}, Vernesa Smol\v{c}i\'c\altaffilmark{7}}

\affil{$^1$Department of Astronomy, Cornell University, Space Sciences
Building, Ithaca, NY 14853, USA \\ $^2$Spitzer Science Center, California Institute of Technology, MC 220-6, 1200 East California Boulevard, Pasadena, CA 91125, USA\\ $^3$National Radio Astronomy Observatory, PO Box O, Socorro,
NM 87801, USA\\ $^4$Cavendish Astrophysics Group, University of Cambridge,
Cambridge, CB3 0HE, UK\\$^5$Argelander-Institut f\"ur Astronomie, Universit\"at Bonn, Auf dem H\"ugel 71, D-53121 Bonn, Germany\\$^6$Astronomy Department, California Institute of Technology, MC 249-17, 1200 East California Boulevard, Pasadena, CA
91125, USA\\ $^7$University of Zagreb, Physics Department, Bijeni\v{c}ka cesta 32, 10002 Zagreb, Croatia\\ }
     \email{$^\dagger$rp462@cornell.edu}
  

\begin{abstract}
We report interferometric measurements of [N{\sc ii}] $205\,{\rm \mu m}$ fine-structure line emission from a representative sample of three galaxies at $z=5$--$6$ using the Atacama Large (sub)Millimeter Array (ALMA). These galaxies were  previously detected in  [C{\sc ii}]  and far-infrared continuum emission and span almost two orders of magnitude in star formation rate (SFR). Our results show at least two different regimes of ionized inter-stellar medium properties for galaxies in the first billion years of cosmic time, separated by their $L_{\rm [CII]}/L_{\rm [NII]}$ ratio. We find extremely low [N{\sc ii}] emission compared to [C{\sc ii}]  ($L_{\rm [CII]}/L_{\rm [NII]}=68^{+200}_{-28}$) from a ``typical"  $\sim L^*_{\rm UV}$ star-forming galaxy, likely directly or indirectly (by its effect on the radiation field) related to low dust abundance and low metallicity. The infrared-luminous modestly star-forming Lyman Break Galaxy (LBG) in our sample is characterized by an ionized-gas fraction ($L_{\rm [CII]}/L_{\rm [NII]}\lesssim20)$ typical of local star-forming galaxies and shows evidence for spatial variations in its ionized-gas fraction across an extended gas reservoir. The extreme SFR, warm and compact dusty starburst AzTEC-3  shows an ionized fraction  higher than expected given its star-formation rate surface density ($L_{\rm [CII]}/L_{\rm [NII]}=22\pm8$) suggesting that [N{\sc ii}] dominantly traces a diffuse ionized medium rather than star-forming H{\sc ii} regions in this type of galaxy. This highest redshift sample of [N{\sc ii}] detections provides some of the first constraints on ionized and neutral gas modeling attempts and on the structure of the inter-stellar medium at $z=5$--6 in ``normal" galaxies and starbursts.
\end{abstract}

\section{Introduction}

The first billion years after the Big Bang is a crucial epoch for understanding galaxy evolution because we can directly witness the initial stages of galaxy assembly. In contrast to present day, most galaxies at these epochs are thought to have only formed a small fraction of their final stellar mass, to be accreting pristine gas very actively from the cosmic web, to have low metal abundances, and to have been affected in their star formation properties by the pristine quality of the inter-stellar medium (ISM) \citep[e.g.,][]{BH07,P09, L10, CN10, RW09, B09, S06}. Investigating the ISM and its relationship to star formation during the first billion years of cosmic time is a promising test bed for galaxy formation models \citep[e.g.,][]{D09a, D09b, D13}, complementing studies of the peak epoch of galaxy assembly ($z\sim2$--$3$; e.g., \citealt{Sh11,CW13,Cas14}).

To faithfully model the interplay of the physical processes at the root of galaxy assembly and evolution we need observations of accurate diagnostics of the different physical phases of the gas. 
While luminous starbursting galaxies like submillimeter galaxies (SMGs) and quasar hosts have been targeted for more than a decade at $z>5$ (e.g., \citealt{M05,W09,R13}; see \citealt{CW13} for a review), we are only now reaching the capability of investigating the ISM in ``normal" star-forming galaxies that are more representative of the general galaxy population at these epochs, in particular Lyman-break galaxies (LBGs; \citealt{R14}, hereafter R14; \citealt{C15}, hereafter C15; \citealt{Wi15}). CO excitation ladders have become a routine tool to investigate the physical conditions in molecular gas for starbursts \citep[e.g.,][]{R10a,R11,R13,We05, Sc11}; however, the prospects for detection in normal galaxies at $z>5$ are not clear due to metallicity effects \citep[][R14]{T13}.  In fact, CO detections of LBGs to date, even exploiting strong lensing, have been limited to $z \lesssim3$ \citep[e.g.,][]{B04,Cop07,R10b,S13}.  


Submillimeter fine-structure lines of the most abundant atomic metal species (mainly C, N, O) and their ions offer a unique angle and an unobscured view of the ISM properties and conditions for star formation that is accessible to ALMA at high redshift. The far-infrared (FIR) cooling lines, less affected by dust attenuation than optical lines, are powerful probes of the star formation activity, linking them directly to the surrounding medium from which stars are born. 
Different lines, better in combination, can be used as diagnostics of the far-ultraviolet (FUV) flux, gas density, temperature, and filling factor of the photon-dominated regions (PDRs) and ionized regions \citep[e.g.,][]{TH85,W90,K06}. Recent surveys with the {\it Herschel Space Observatory} provide the context for high redshift studies by giving benchmark
sets of local galaxies, for which almost complete suites of FIR lines yield solid constraints and allow detailed ISM modeling \citep[e.g.,][]{Ro15,Co15,DL14,S15,K16}.


The ionization potential of nitrogen ($14.5\,$eV) is greater than that of hydrogen, therefore the singly ionized nitrogen lines, [N{\sc ii}], probe the effect of UV photons emitted by massive young stars. Since N$^+$ is only present in the ionized medium it is a tracer of the extended low density envelope of H{\sc ii} regions, the ionized surfaces of dense atomic and molecular clouds and the warm ionized medium (WIM). \cite{GS15} find that the dominant source of [N{\sc ii}] emission in the Milky Way is not the WIM, because the electron density they measure from [N{\sc ii}] fine-structure line ratios is two orders of magnitude higher than what is expected in the Galactic diffuse medium, but it is expected that different types of galaxies, at different ages, will differ significantly in their ISM phase structure.

The high brightness of the [C{\sc ii}] $158\,\mu$m line makes it an ideal target at high redshift, where it allows dynamical studies and a probe into the ISM properties and star formation. However, [C{\sc ii}] can originate from a range of gas phases and the line luminosity alone gives no direct information on its origin. In fact, low-$z$ observations have shown that while most of the [C{\sc ii}] luminosity comes from the atomic PDRs and diffuse cold neutral medium (CNM), significant fractions also come from ionized gas regions and CO-dark molecular clouds \citep{P13}. 
\cite{O06,O11} suggested that the [N{\sc ii}] $205\,\mu$m line could be used in conjunction with  [C{\sc ii}] $158\,\mu$m  to separate the ionized from neutral fraction of the ISM, because the latter is emitted by both weakly ionized and atomic gas.  The transition critical densities of these two lines in ionized gas are very similar, implying that the line ratio for the ionized medium is nearly constant ($\sim3$--4, \citealt{O06}) and set by the relative carbon to nitrogen abundance, making [N{\sc ii}] a tracer of the ionized fraction of [C{\sc ii}]-emitting gas.

\cite{L16} combined [N{\sc ii}] and [C{\sc ii}] Galactic observations to determine the fraction of ionized gas contributing to the [C{\sc ii}] emission in the Milky Way for many lines of sight, finding this to vary mostly between 0.5 and 0.8 (corresponding to [C{\sc ii}]/[N{\sc ii}] ratios $\sim$4--10), where the Galactic center line of sights have the smallest ionized fraction, probably due to higher gas densities.


The [N{\sc ii}] $205\,\mu$m line is the tool of choice to characterize the low-density ionized ISM at high redshift. Thus, we here use ALMA to measure the [N{\sc ii}]  properties of a sample spanning the variety of star-forming activity known at $z=5$--6: a hyper-luminous nuclear starburst (AzTEC-3; SFR$\,{\rm\sim1100\,M_{\odot}\,yr^{-1}}$), a dusty, high star formation rate LBG (HZ10; SFR$\,{\rm\sim170\,M_{\odot}\,yr^{-1}}$) and a typical, less star-forming LBG (LBG-1; SFR$\,{\rm\sim10-30\,M_{\odot}\,yr^{-1}}$). These last two galaxies are near $L^*_{\rm UV}$ for LBGs at $z=5$--6, and can be considered ``typical" because they are consistent with the current constraints on the star-forming main sequence of galaxies at these redshifts \citep[e.g.,][]{Sp14}. All galaxies were previously detected in [C{\sc ii}] with ALMA (R14; C15).

The structure of this paper is as follows. In Section~\ref{obs} we describe the ALMA observations of the [N{\sc ii}] and [C{\sc ii}] lines utilized in this study. Our results are presented in Section~\ref{res}, and in Section~\ref{model} we describe the H{\sc ii} and PDR models we have utilized to assist our interpretation. In Section~\ref{analysis}, we present our interpretation for each galaxy in our sample. We close with a discussion of the impact of our findings and of this kind of FIR line studies in the first giga-year of cosmic time in Section~\ref{discuss}. 
In this work we assume a flat $\Lambda$CDM cosmology with $H_0$=$70\,$km s$^{-1}$ Mpc$^{-1}$, $\Omega_{\rm m}$=0.3 and  $\Omega_\Lambda$=0.7.

\section{Observations}
\label{obs}
\subsection{ALMA Cycle-3 observations of {\rm [N{\sc ii}]}}
We observed the [N{\sc ii}] $205\,\mu$m ($^3P_1\to~^3P_0$) transition line ($\nu_{\rm rest}$=1461.131 GHz, redshifted to $\sim$220--$230\,{\rm GHz}$ at $z\sim5.3$--5.7), using ALMA in Band 6 (see Table~\ref{table_lines} for details). Observations were carried out in Cycle 3, with 32--47 usable $12\,$m antennae under good weather conditions at 1.3 mm (precipitable water vapor columns of $2.8\,$mm for AzTEC-3, $1.1\,$mm for HZ10 and 1.6--$1.8\,$mm for LBG-1) for 4 tracks between 2015 December 16 and 2016 January 14, using 3 tracks in a compact configuration (max. baseline $\sim300\,$m) and one track on LBG-1 in transition from an extended to a compact configuration (min. baseline $15\,$m, max. baseline $6\,$km; two additional tracks were discarded due to poor data quality because of weather).
These observations resulted in total on-source times of $25\,$min for AzTEC-3, $50\,$min for HZ10 and $64\,$min for LBG-1, with standard calibration overheads.

The nearby radio quasar J0948+0022 was observed regularly for amplitude and phase gain calibration, J1058+0133 and J0854+2006 were observed for bandpass and flux calibration, and Ganymede was also used for flux calibration of two tracks. We estimate the overall accuracy of the flux calibration to be within $\sim10\%$.

For each target the correlator was set up to cover two spectral windows of 1.875 GHz bandwidth each at $3.9\,$MHz ($\sim5\,$km s$^{-1}$) resolution (dual polarization) in each sideband.
The [N{\sc ii}] line was centered in one of the spectral windows, in the upper sideband for AzTEC-3 and LBG-1 (also covering the CO($J$=12$\to$11) line in the lower sideband) and in the lower sideband for HZ10, and the other three spectral windows were placed to provide contiguous coverage of the continuum emission within the available bandwidth.

The Common Astronomy Software Application (CASA) version 4.5 was used for data reduction and analysis. All images and mosaics were produced with the \rm{CLEAN} algorithm, using natural weighting for maximal sensitivity. We also apply an outer taper of $300\,$k$\lambda$ to the LBG-1 observations that included unwanted long baselines ($>300\,$m).

AzTEC-3 and LBG-1 are separated by only $15^{\prime\prime}$, so their individual pointings were imaged both separately and together in a mosaic. This significantly improves our sensitivity for AzTEC-3 (primary beam correction factor of $\sim$0.7 in the mosaic), due to the longer on-source time of the LBG-1 pointing. We therefore extracted the AzTEC-3 line and continuum fluxes from the mosaic, while LBG-1 was analyzed in the single pointing, after confirming that the results were equivalent in both cases.

The mosaic has a synthesized beam size of $1.4^{\prime\prime} \times 1.2^{\prime\prime}$ and the rms noise is $\sim0.34\,$mJy beam$^{-1}$ at the position of AzTEC-3 in $41\,$km s$^{-1}$ wide channels and $\sim0.21\,$mJy beam$^{-1}$ at the position of LBG-1. The rms noise in the continuum maps are $\sim 20\,\mu$Jy beam$^{-1}$  and $\sim 15\,\mu$Jy beam$^{-1}$  at the positions of AzTEC-3 and LBG-1 in the mosaic, respectively.
The LBG-1 pointing has a synthesized beam of $1.3^{\prime\prime} \times 1.1^{\prime\prime}$ and the rms noise is also $\sim0.21\,$mJy beam$^{-1}$ in the middle of the field in $41\,$km s$^{-1}$ channels, and the continuum rms noise is $\sim 16\,\mu$Jy beam$^{-1}$ .

For HZ10, the imaging results in a synthesized beam size of $1.7^{\prime\prime} \times 1.2^{\prime\prime}$ at the redshifted  [N{\sc ii}] frequency and in the continuum map. The rms noise in the phase center is $\sim0.2\,$mJy beam$^{-1}$ in a $44\,$km s$^{-1}$ wide channel and the final rms noise when averaging over all spectral windows (i.e. over a total $7.5\,$GHz of bandwidth) is $\sim 35\, \mu$Jy beam$^{-1}$.

\subsection{Archival ALMA Cycles 0 \& 1 observations of {\rm [C{\sc ii}]}}
The ALMA data targeting the [C{\sc ii}] lines for our sample galaxies have previously been presented by R14 for the Cycle-0 observations of AzTEC-3 and LBG-1, and by C15 for the Cycle-1 observations of HZ10 and LBG-1 (referred to as HZ6 by C15).
The Cycle-1 observations were taken on 2013 November 15-16 in band 7 as part of a larger project; we here provide a brief description of the data that we have re-analyzed from C15.
The HZ10 pointing resulted in $56\,$min on source with 25 usable antennae. Ganymede was observed as flux calibrator, J0538--4405 was observed as bandpass calibrator, and J1058+0133 was observed as amplitude/phase gain calibrator.
The LBG-1 (HZ6) data taken in Cycle-1 resulted in $42\,$min on source with 29 antennae. Pallas was observed as flux calibrator, and the bandpass and gain calibrators were the same as for the HZ10 observations.
In both cases the correlator was set up to target the expected frequency of the [C{\sc ii}] line and to provide continuous coverage of the continuum emission in adjacent spectral windows with channels of $15.6\,$MHz in Time Division Mode (TDM).

We have re-calibrated these archival data and the Cycle-0 data described by R14 using the scripted pipelines on the archive after appropriately modifying them to be executed with CASA 4.5 and using the updated flux calibration tables from Butler-JPL-Horizons 2012 for Solar System objects. This modification lowers the fluxes of the Cycle-0 data compared to R14 by $\sim15\%-20\%$ because of a corresponding lowering of the assumed flux of Callisto compared to Butler-JPL-Horizons 2010.
We have furthermore modified the default flags at the edges of the spectral windows that cover the [C{\sc ii}] line in HZ10, to only flag two channels per window. This is sufficient because the bandpass calibration shows a sharp drop-off, and by imaging both spectral windows we recover the full line emission.

We have produced image cubes and continuum maps using the {\rm CLEAN} task with natural weighting in CASA 4.5. LBG-1 was imaged by mosaicking  Cycle-0 and Cycle-1 data as part of the {\rm CLEAN} process.  AzTEC-3 was imaged separately using the Cycle-0 data for the single pointing close to the source, after checking that mosaicking the second pointing (centered near LBG-1) does not significantly alter the results due to the high signal-to-noise ratio of the continuum and line detection in this source.

\begin{figure*}[tbh]
\hspace{-0pt}\includegraphics[scale=.72]{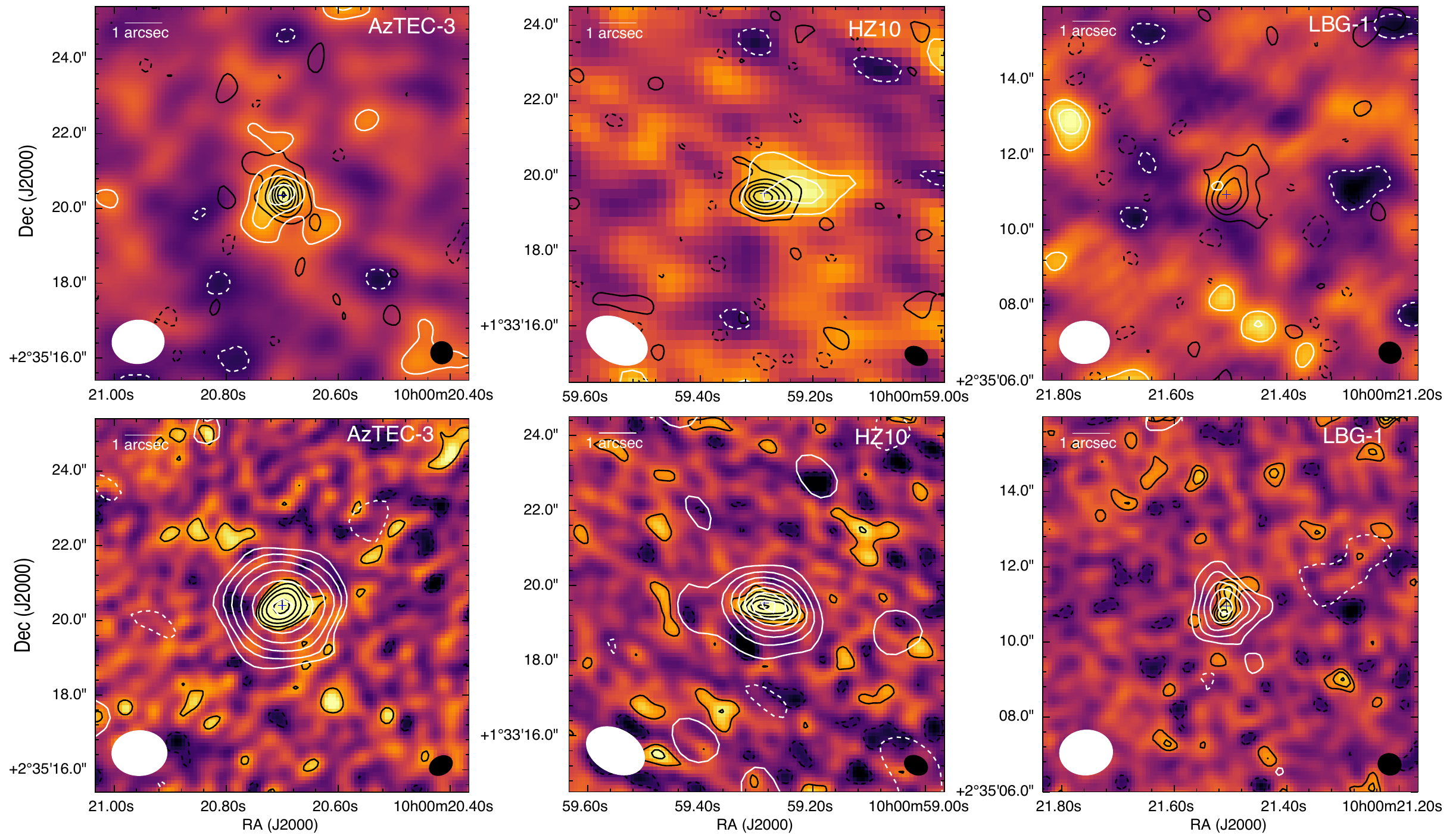}
\caption{Top: Integrated line maps (over the line FWHM) showing [N{\sc ii}] color-scale with  [N{\sc ii}] (white) and [C{\sc ii}] (black) contours. Blue crosses indicate the positions of the $205\,\mu$m continuum peak. The  [N{\sc ii}] ([C{\sc ii}]) beam is shown in the bottom left (right) corner of each panel. The  [N{\sc ii}] ([C{\sc ii}]) contours are multiples of $1\sigma$ ($4\sigma$), starting at $\pm2\sigma$. The  [N{\sc ii}] in the starburst AzTEC-3 is a 5$\sigma$ detection, it is consistent with being centered at the position of the [C{\sc ii}] and continuum emission, and it may be resolved.  In the FIR-bright LBG, HZ10, the  [N{\sc ii}] emission (3.2$\sigma$ per beam at the peak) is slightly resolved (5.3$\sigma$ integrated significance) and appears to be offset with respect to the center of the [C{\sc ii}]-emission, indicating significant variation of [C{\sc ii}]/[N{\sc ii}] across the galaxy. The significance of the  tentative [N{\sc ii}]  detection in LBG-1 is only $2\sigma$, which constrains the [C{\sc ii}]/[N{\sc ii}] ratio to be very high ($68^{+200}_{-28}$). 
Bottom: Continuum maps showing $158\,\mu$m color-scale with $205\,\mu$m (white) and $158\,\mu$m (black) contours.  Contours start at $\pm2\sigma$ and are at powers of $2\sigma$ (i.e. $\pm2\sigma,4\sigma,8\sigma...$) for AzTEC-3;  in steps of $2\sigma$ for HZ10 and in steps of $1\sigma$ for LBG-1. The $205\,\mu$m ($158\,\mu$m) beam is shown in the bottom left (right) corner.}
\label{mom0}
\end{figure*}

\begin{figure*}[tbh]
\hspace{-15pt}\includegraphics[scale=0.33]{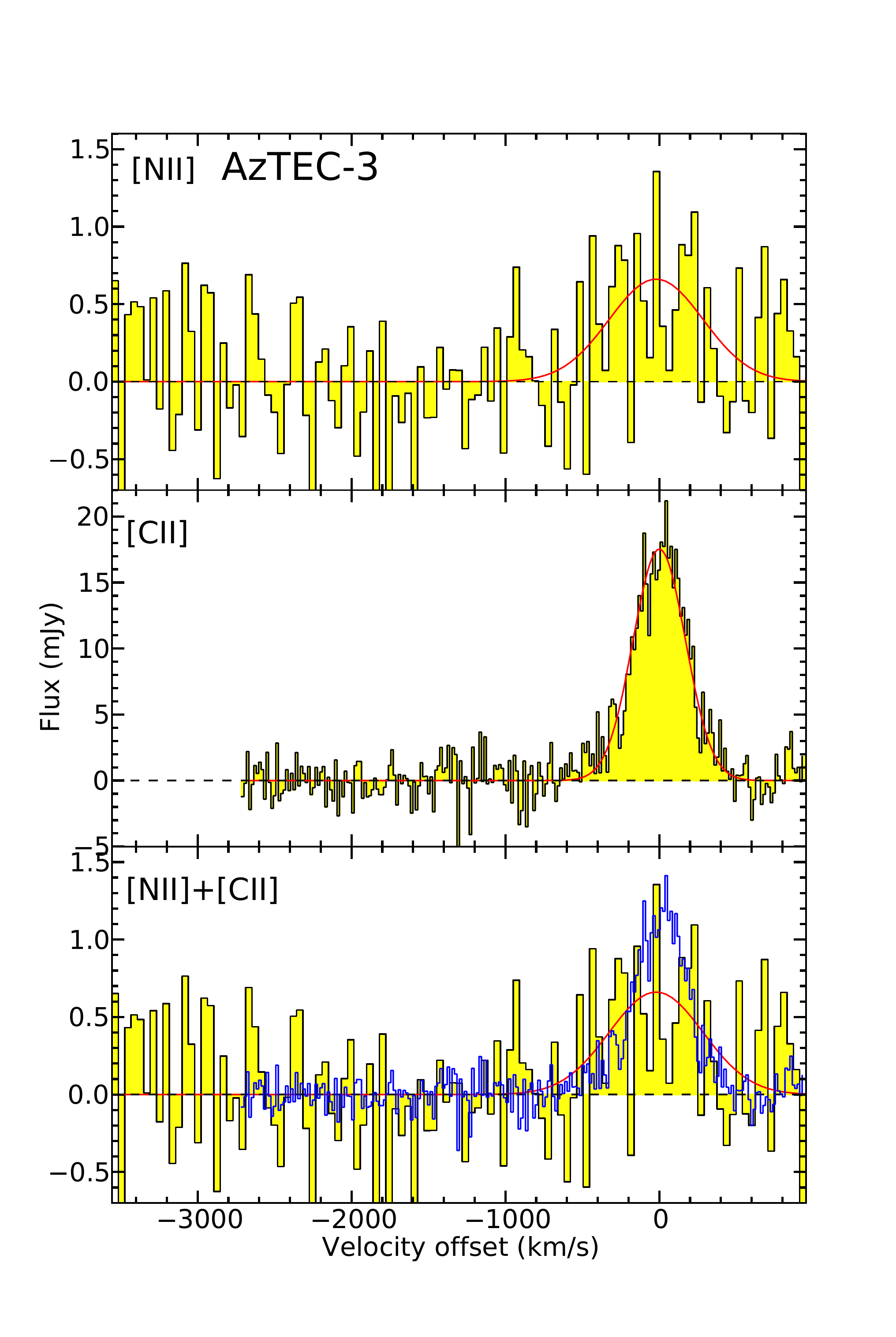}\hspace{-15pt}\includegraphics[scale=0.33]{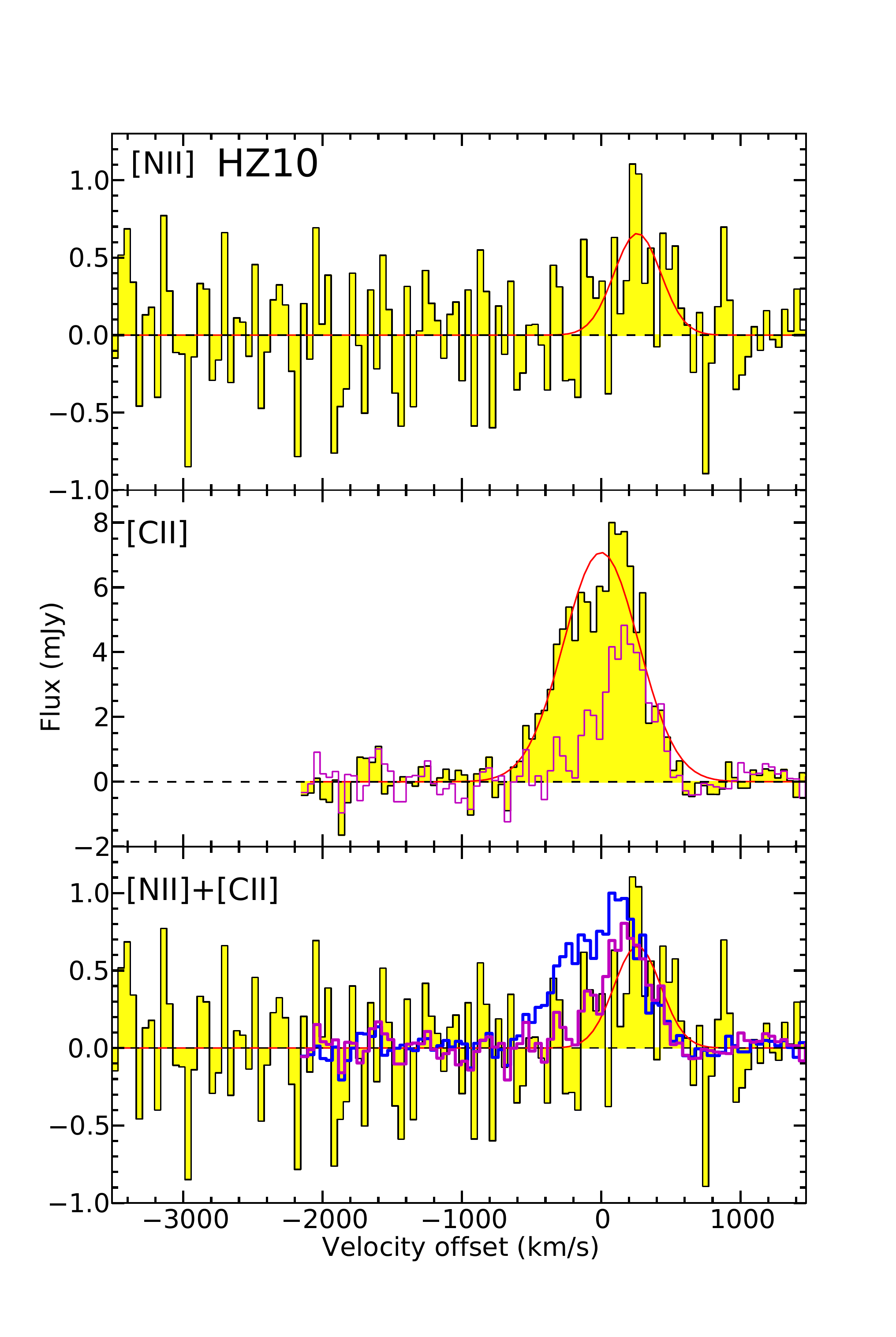}\hspace{-15pt}\includegraphics[scale=0.33]{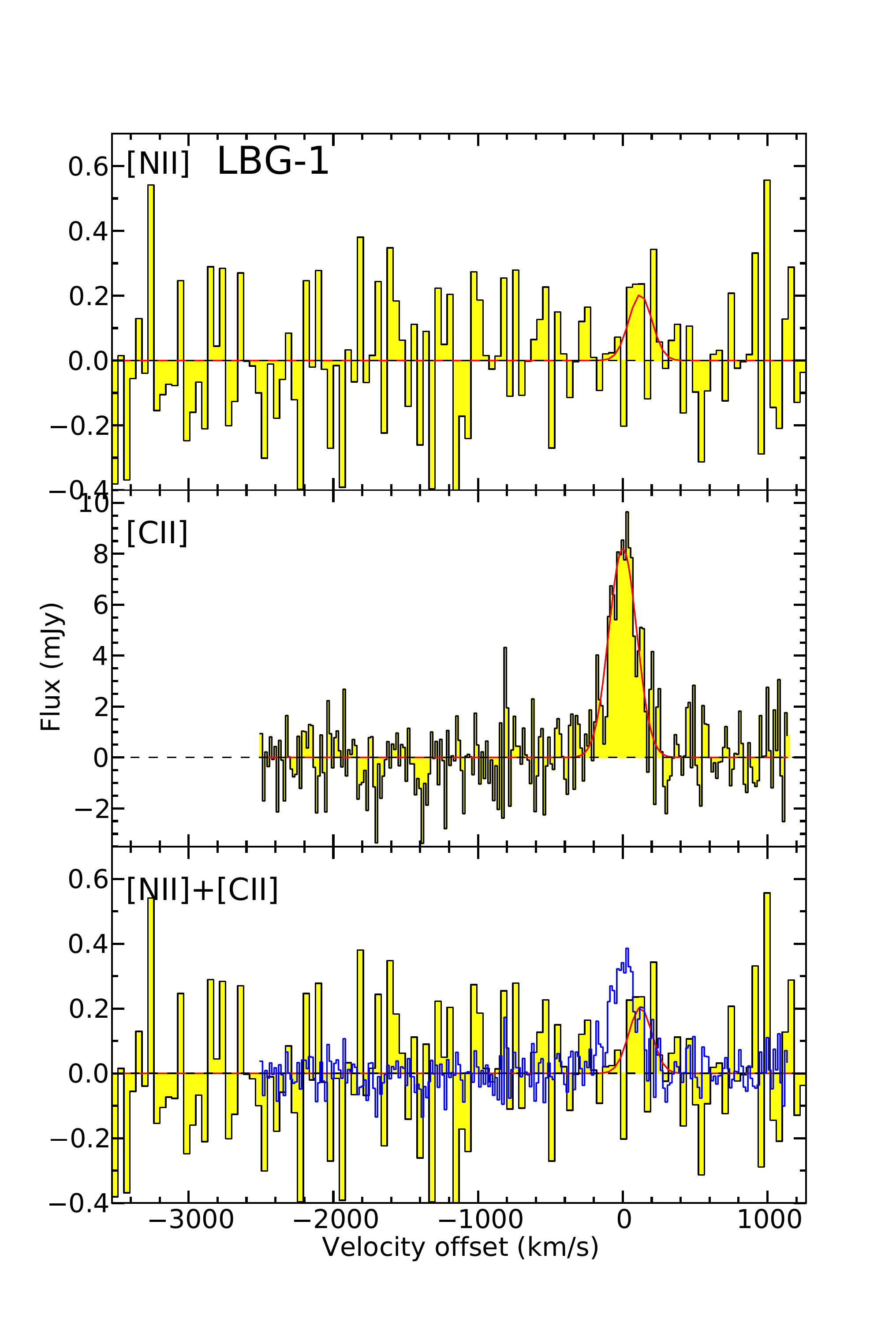}
\caption{[C{\sc ii}] and [N{\sc ii}] spectra of our sample galaxies at $z=5$--6, and Gaussian fits to the line emission (red curves). The channel velocity width in all [N{\sc ii}] spectra is $\sim42\,$km s$^{-1}$; in the [C{\sc ii}] spectra it is  $\sim16\,$km s$^{-1}$ in AzTEC-3 and LBG-1 and $\sim44\,$km s$^{-1}$ in HZ10. [C{\sc ii}] is scaled arbitrarily down in flux density in the bottom panels for comparison.  The [N{\sc ii}] line appears to be broader than [C{\sc ii}] in AzTEC-3. HZ10 shows a velocity shift between [C{\sc ii}] and [N{\sc ii}]. The magenta spectra in the central middle and bottom panels show a single-pixel [C{\sc ii}] spectrum extracted at the [N{\sc ii}] peak position (``map peak" method, as defined in Section~\ref{resHZ10}; in units of mJy\,beam$^{-1}$, after smoothing to the [N{\sc ii}] resolution), suggesting a common kinematic origin with [N{\sc ii}]. }\label{spectra}
\end{figure*}

\section{Results}
\label{res}

\subsection{The massive dusty starburst AzTEC-3}

We detect the [N{\sc ii}] line (Table~\ref{table_lines}) in AzTEC-3 at 5$\sigma$ confidence in the mosaic of the AzTEC-3 pointing and the LBG-1 pointing (the significance is 3.5$\sigma$ in the AzTEC-3 data alone).
The integrated line map (Fig.~\ref{mom0}) shows tentative evidence ($\sim3\sigma$) for extended structure in the [N{\sc ii}] emission, over a scale of $(2.8^{\prime\prime} \pm1.0^{\prime\prime} ) \times (0.7^{\prime\prime} \pm 0.6^{\prime\prime} )$ P.A. 15 degrees (deconvolved).

Because of the moderate signal-to-noise ratio and consequently poorly constrained spatial extent of the emission we extract the  [N{\sc ii}] line spectrum using an aperture of $1.5^{\prime\prime} \times1.3^{\prime\prime} $, which is fitted to the size of the [C{\sc ii}] emission convolved with the [N{\sc ii}] beam. This yields a line peak flux in the spectrum of $(0.65\pm0.14)\,$mJy.\footnote{Unless otherwise stated, all aperture spectra are extracted in apertures fitted to the FWHM of the spatial emission, and then corrected by a factor of $\times 2$, to account for the missed flux outside the aperture, based on the assumption of Gaussian spatial profiles.} The flux measured through 2D-Gaussian fitting in the integrated line map would result in [N{\sc ii}] line flux which is $\sim$25\% higher, compatible with our measurement to within $\sim1\sigma$. However, given the significance of the detection of only $\sim5\sigma$, the spatial size of the Gaussian fit is highly uncertain. This leads us to adopt the [C{\sc ii}]-based aperture flux in the following.

The measured width of the [N{\sc ii}] line profile appears $\sim60\% \pm40\%$ greater than [C{\sc ii}], which could be indicative of a different origin for the gas traced by [N{\sc ii}] and the [C{\sc ii}] emission, however the significance of this difference is only 1.4$\sigma$.
By fitting the FWHM of the [N{\sc ii}] line (Fig.~\ref{spectra}, $660\pm180\,$km s$^{-1}$) we measure an integrated line flux of $(0.46\pm0.16)\,$Jy km s$^{-1}$, yielding a [C{\sc ii}] to [N{\sc ii}] luminosity line ratio of 22$\pm$8. 
Due to the poorly constrained velocity width of the line it is worth considering the possibility that the real FWHM of [N{\sc ii}] be the same as in [C{\sc ii}]. Fixing the [N{\sc ii}]  line width to the [C{\sc ii}]  line width in the fitting would yield a ratio of $42\pm9$.



\subsection{The FIR-luminous LBG HZ10}
\label{resHZ10}
We detect the [N{\sc ii}] line in HZ10 with an integrated significance of $\sim 5.3\sigma$ (peak signal-to-noise ratio of $\sim$3.2 per beam).

The [C{\sc ii}] emission is spatially resolved and shows a smooth velocity gradient suggestive of a disk, with deconvolved size of $(0.80^{\prime\prime} \pm 0.07^{\prime\prime}) \times (0.42^{\prime\prime} \pm 0.06^{\prime\prime})$. The [N{\sc ii}] emission also appears to be extended with a marginally larger size of $(1.9^{\prime\prime} \pm 0.9^{\prime\prime}) \times (0.6^{\prime\prime} \pm 0.7^{\prime\prime})$, and it appears to be associated with only one part of the [C{\sc ii}] emitting region (offset $\sim0.8^{\prime\prime} \pm 0.3^{\prime\prime}$). 
Figs.~\ref{mom0} and \ref{spectra} show that both the velocity and position shift relative to the [C{\sc ii}] emission are compatible with the interpretation of differential emission, with [N{\sc ii}] coming predominantly from a location offset spatially and kinematically with respect to the center of the [C{\sc ii}] and FIR emission.
Because the [N{\sc ii}] emission appears extended and not coincident with the [C{\sc ii}], we extract the [N{\sc ii}] line spectrum in an aperture of size equal to the FWHM of the 2D-Gaussian fit to the [N{\sc ii}] integrated line map.


The global [C{\sc ii}] to [N{\sc ii}] luminosity ratio obtained from the integrated line fluxes is $19\pm6$ for HZ10, but this does not take into account the potentially different origin of the emission. In order to investigate the spatial and kinematic variations of the line ratio across the galaxy, we explore two alternative methods to estimate the amount of [C{\sc ii}] luminosity co-spatial with the [N{\sc ii}] emission.
In the first method (``red") we create a moment-0 map of the [C{\sc ii}] line emission over the FWHM velocity range from the Gaussian fit to the spectrum of the [N{\sc ii}] line emission to extract the fraction of the [C{\sc ii}] line flux associated with the [N{\sc ii}] emission.  This is motivated by the finding that the detected [N{\sc ii}] only appears to be emitted from the red part of the [C{\sc ii}] velocity range. This shifts the center of the  [C{\sc ii}] peak by $0.2^{\prime\prime}$ in the direction of the [N{\sc ii}] peak and gives a [C{\sc ii}]/[N{\sc ii}] line luminosity ratio of $14\pm5$, for the red part of HZ10 (``HZ10-red" in Fig.~\ref{ratio_plot}).
In the second method (``map peak") we measure the [C{\sc ii}] line flux in a single pixel at the position of the [N{\sc ii}] peak (after convolving the [C{\sc ii}] data-cube to the resolution of the [N{\sc ii}]  map). Due to the fact that the [N{\sc ii}] may be extended, this method yields a lower bound on the [C{\sc ii}] flux from the [N{\sc ii}]-emitting region. This provides a constraint on the [C{\sc ii}]/[N{\sc ii}] ratio in the region of the galaxy where the [N{\sc ii}] emission is strongest, and yields a low ratio of $8\pm3$. 

\subsection{The ``normal" $L^*$ galaxy LBG-1}

Our sensitive data on LBG-1 only resulted in a tentative ($2\sigma$) detection of the [N{\sc ii}] line at $(0.2\pm0.1)\,$mJy peak flux. The emission appears unresolved and its position and velocity width are compatible with that of the [C{\sc ii}] line within the uncertainties. The single-pixel flux we extract at the peak of the integrated line map is compatible with the aperture flux, within the significant uncertainties. There is tentative evidence for a slight velocity redshift of the [N{\sc ii}] line with respect to the [C{\sc ii}] by $(120\pm40)\,$km s$^{-1}$.\footnote{Determined through $\chi^2$ fitting of a Gaussian line profile without imposing any priors on the line position or width.}
Although the moderate signal-to-noise does not allow us to conclusively establish this offset, it appears to be reminiscent of our findings in HZ10.
The extremely low value of the [N{\sc ii}] luminosity constrains the ratio of [C{\sc ii}] to [N{\sc ii}] line luminosities to be $\sim68^{+200}_{-28}$, which may indicate ISM properties never seen before at high redshift.


\subsection{Continuum measurements}
We extract continuum fluxes and sizes from 2D-Gaussian fitting or aperture fluxes for each source both in the archival $158\,\mu$m and our new $205\,\mu$m data. All the sources are detected with moderate to high significance. Table~\ref{table_cont} presents the results of the flux measurements in the two separate sidebands, where the signal-to-noise ratio is sufficient.

In order to measure the far-infrared luminosity, we explore the parameter space of modified black-body models with a Markov chain Monte Carlo (MCMC) routine to fit the continuum measurements for the LBGs (HZ10 and LBG-1) whose FIR SED shapes were previously unconstrained.\footnote{The FIR SED of AzTEC-3 was already constrained by numerous previous studies and is not significantly improved by the current measurement.}
The FIR luminosity implied by our fits is strongly dependent on the assumed dust temperature, which is poorly  constrained by our measurements; since the dust emissivity index $\beta$ and the dust temperature are degenerate on the Rayleigh-Jeans tail of the dust emission. We assume optically thin FIR emission and impose weak Gaussian priors on the dust temperature ($30\pm20\,$K) and $\beta$ ($1.7\pm0.5$) based on the SEDs of local dwarf galaxies \citep{RR13} and compatible with Milky Way values, to obtain dust temperatures of $33^{+12}_{-8}\,$K for HZ10 and  $38^{+16}_{-13}\,$K for LBG-1. 
The resulting FIR luminosities (between 42.5 and $122.5\,\mu$m) implied by the fits are $6.5^{+9.1}_{-3.8} \times 10^{11} \,L_\odot$ for HZ10 and $1.8^{+3.3}_{-1.3} \times 10^{11}\, L_\odot$ for LBG-1, compatible with previous estimates (C15).
If we relax the prior on the dust temperature to a uniform prior between 10--$100\,$K, we obtain
$36^{+25}_{-10}\,$K for HZ10  and $60^{+35}_{-27}\,$K for LBG-1, respectively, showing that the dust temperature is significantly uncertain given current constraints, and potentially allowing for higher FIR luminosity, if the dust temperature were higher than expected. In fact, the remaining uncertainty on the upper end of the dust temperature probability distribution, and thus the FIR luminosity, implies that ``normal" systems like LBG-1 may have higher than previously estimated FIR luminosity (C15), and thus may remain consistent with the SMC IRX--$\beta_{\rm UV}$ relation between the infrared and ultraviolet luminosity ratio and the ultraviolet continuum slope discussed in C15, despite apparently falling below when ``average" dust temperatures are assumed.

\begin{figure}[tbh]
\hspace{-15pt}\includegraphics[scale=0.5]{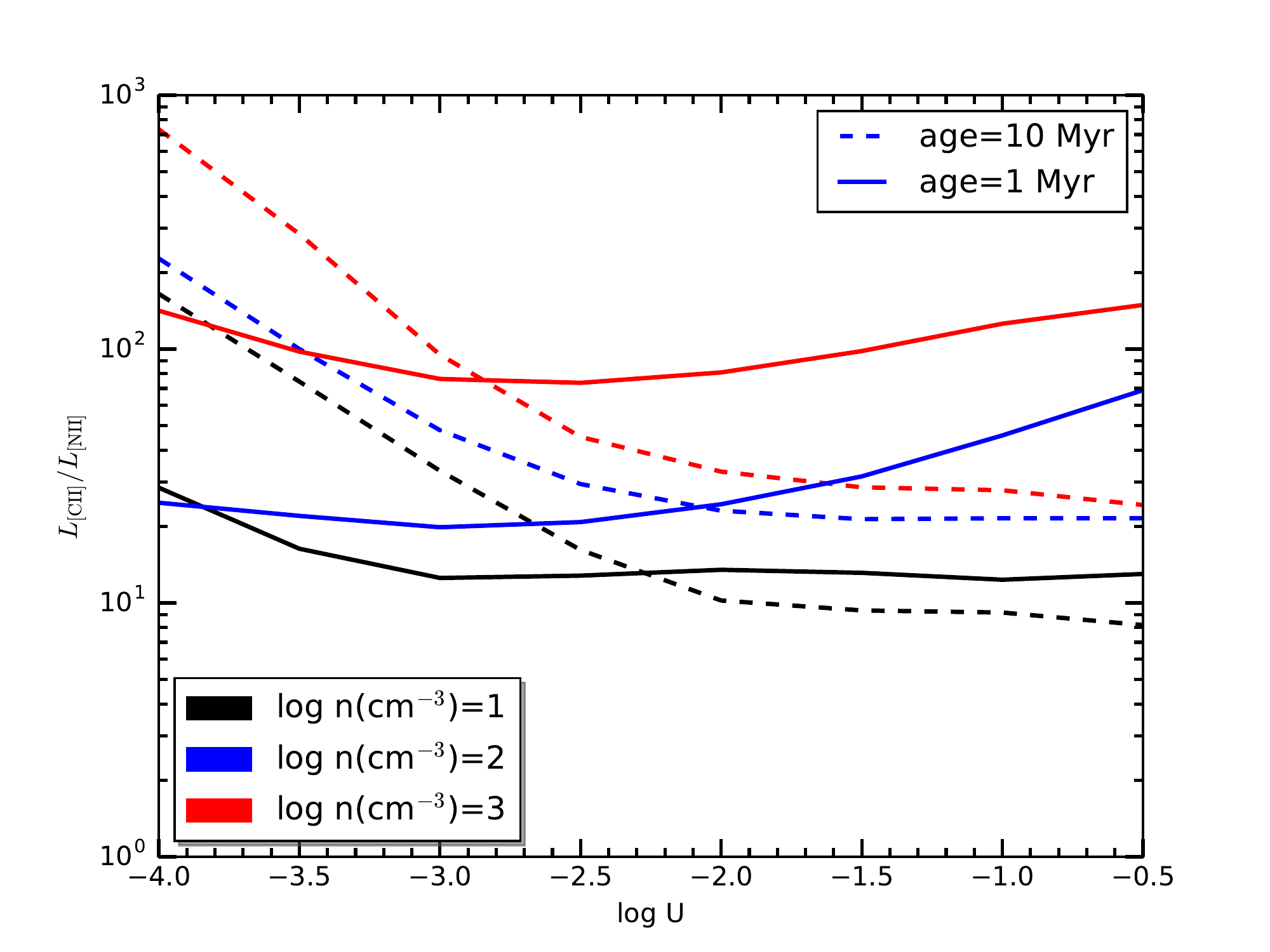}
\caption{Cloudy model results for the ratio of the [C{\sc ii}]  and [N{\sc ii}]  line luminosities for a grid of H{\sc ii} region+PDR models. The age of the starburst (1 and 10 Myr, solid and dashed lines) determines the mass of the most massive Main-Sequence star still present and affecting the hardness of the radiation field. The main difference between the different age tracks is that the 1 Myr starburst produces a significant $\rm N^{++}$ region, while the 10 Myr starburst does not. The density range (from $10\,$cm$^{-3}$ to $1000\,$cm$^{-3}$) covers diffuse ionized gas to ultra-compact H{\sc ii} regions. The substantial uncertainty on the intensity of the radiation field at high redshift, in very different environments,  is captured by the broad range of ionization parameter (U) considered, along the abscissa. Of particular note are the strong density effect (high density implies higher line ratio) and the slight rise of the ratio for hard radiation and high intensities due to further ionization of N$^+$ to N$^{++}$.}
\label{Cloudy}
\end{figure}

\begin{table*}[t]
\caption[]{Measured [C{\sc ii}] and [N{\sc ii}] line properties of our sample galaxies}
\label{table_lines}
\resizebox{2\columnwidth}{!}{%
\begin{tabular}{l c c c c c c c}
\hline 
Quantity & AzTEC-3 & HZ10 & LBG-1\\
\hline \noalign {\smallskip}
[C{\sc ii}] line properties\\
$\nu_{{\rm obs}}$(GHz) &$301.771\pm0.006$ & $285.612 \pm 0.013$ &  $301.980\pm0.007$ \\

Redshift&$5.29795\pm0.00013$&$5.6543 \pm 0.0003$&$5.29359\pm0.00015$ \\
$S_{{\rm [CII]}}$(mJy) &  $17.5 \pm 0.7 $ &  $7.1 \pm0.3$ &  $8.2\pm0.5$ \\
$FWHM_{\rm [CII]}$ (km s$^{-1}$) & $410 \pm 15$ &  $630 \pm 30$ &$230\pm20$  \\
$I_{\rm [CII]}$ (Jy km s$^{-1}$) & $7.8\pm 0.4$ & $4.5 \pm 0.3$ &$2.1\pm0.2$ \\
$L_{\rm [CII]}$ ($10^{9} L_\odot$)&$6.4\pm0.3$& $4.0\pm0.3$&$1.71\pm0.16$\\
deconvolved size & $(0.65^{\prime\prime}\pm0.06^{\prime\prime}) \times (0.33^{\prime\prime}\pm0.07^{\prime\prime})$  & $(0.80^{\prime\prime}\pm0.07^{\prime\prime}) \times (0.42^{\prime\prime}\pm0.06^{\prime\prime})$ &$(1.00^{\prime\prime}\pm0.12^{\prime\prime}) \times (0.57^{\prime\prime}\pm0.10^{\prime\prime})$ \\ 
size (kpc$^2$) &$(4.0\pm0.4)\times  (2.1\pm 0.4)$&$(4.8\pm 0.4)\times (2.5\pm0.4)$& $(6.2\pm 0.7)\times  (3.5\pm0.6)$\\
\hline
[N{\sc ii}] line properties\\
$\nu_{{\rm obs}}$(GHz) & $232.02\pm0.06$ & $219.39 \pm 0.04$ & $232.1\pm0.3$  \\
$S_{{\rm [NII]}}$ (mJy) & $0.65\pm 0.14$ & $0.66 \pm0.18$ &  $0.20 \pm 0.09$ \\
$FWHM_{\rm [NII]}$ (km s$^{-1}$) & $660\pm180$ & $400 \pm 120$ & $190 \pm 90$ \\
$I_{\rm [NII]}$ (Jy km s$^{-1}$) &$0.46\pm0.16$&  $0.31 \pm 0.11$  & $0.04\pm0.03$ \\
$L_{\rm [NII]}$ ($10^{9} L_\odot$)&$0.29\pm0.10$&$0.21\pm0.08$&$0.025\pm0.019$\\
\hline
$L_{\rm [CII]}/L_{\rm [NII]}$ & $22\pm8$& $19\pm6$ (``red": $14\pm5$) &$68^{+200}_{-28}$ \\
\hline \noalign {\smallskip}
\end{tabular}
}
\end{table*}

\begin{table}[t]
\caption[]{Measured continuum properties of our sample galaxies}
\label{table_cont}
\resizebox{1\columnwidth}{!}{%
\begin{tabular}{ c c c | c c }
\hline 
Target &  $\nu_{\rm obs}$ (GHz) & $S_\nu$ (mJy) & $\nu_{\rm obs}$ (GHz) & $S_\nu$ (mJy)\\
\hline \noalign {\smallskip}
AzTEC-3 & 303.62 &  $5.2 \pm 0.3$ & 233.93 & $3.19\pm0.09$ \\
 & 289.87& $4.4 \pm 0.3$  & 217.61 & $2.58\pm 0.09$ \\
HZ10 &   297.84&$1.21\pm 0.24$  &234.50& $0.87 \pm 0.14$ \\
 &   285.81&$1.05\pm 0.20$ & 220.37& $0.69 \pm 0.11$  \\
LBG-1 & 296.79& $0.22 \pm 0.07$& 233.06&$0.14 \pm 0.04$  \\
 & & & 218.62&$0.11 \pm 0.04$  \\
\hline \noalign {\smallskip}
\end{tabular}
}
\end{table}


\newpage

\section{Cloudy modeling of the {\rm [C{\sc ii}]/[N{\sc ii}]} ratio}
\label{model}
In order to aid the interpretation of our observations we have run a grid of Cloudy models using version 13.03 \citep{cl}. We simulate constant pressure clouds with spherical geometry (although the inner radius of $3\,$kpc that we adopt implies the H{\sc ii} region and PDR are close to the plane-parallel regime), extending up to $A_V$=10 to include both H{\sc ii} regions and PDRs. The input spectrum is taken from Starburst99 models \citep{stb99} with Geneva stellar tracks \citep{Geneva}, default parameters for solar metallicity stellar atmospheres and a burst of star formation in order to use the age of the starburst as a proxy for hardness of the radiation field. We also explore the effect of low stellar metallicity tracks ($Z=0.001$, i.e. $\sim 0.07\,Z_\odot$) on our results but find little difference for the main quantities of interest.
We adopt all gas element abundances for solar metallicity from \cite{N11} and include Orion-like dust and magnetic fields (intensity of $10^{-5}\,$G at the inner edge, as appropriate for the Orion nebula according to the Cloudy documentation). We fix the cloud inner edge gas density and ionization parameter (U) for starburst ages of 1 Myr and 10 Myr, and vary both parameters in a grid.
Figure~\ref{Cloudy} shows the integrated line luminosities predicted by our Cloudy simulations from large, single clouds illuminated by a central starburst. While the precise values are model-dependent and should not be directly compared to the measured values (and depend on the precise carbon to nitrogen abundance ratio and gas density structure) we examine the main trends to gain insight into the major factors affecting the [C{\sc ii}]/[N{\sc ii}] line luminosity ratio.

The main trend is a clear density dependence, with higher density (for a fixed ionization parameter, U) corresponding to higher $L_{\rm [CII]}$/$L_{\rm [NII]}$. 
Although this effect does not appear straightforward to interpret, we suggest some elements that are likely to play a role.
The effect that we see in these models is probably partly caused by a saturation of the [N{\sc ii}] total emission with increasing density, i.e., a lack of substantial increase due to the low critical density ($\sim 100\,$cm$^{-3}$) of the [N{\sc ii}] transition. At the same time the Cloudy models show a steady increase of the [C{\sc ii}]  emission from the PDR with the simultaneous increase in density and radiation field (an increase in density at constant U implies a proportional increase of the input stellar flux).
The gas density in the PDR is higher than in the H{\sc ii} region but not as high as one might expect from simple gas pressure balance; this is due to the magnetic pressure becoming important in the PDR for some of the low and intermediate density models. As a consequence the density in the PDR approaches the [C{\sc ii}] critical density for neutral gas from below and only crosses it for the high density models, with [C{\sc ii}] saturating at higher H{\sc ii} gas densities than naively expected.
Thus, the increase of the gas density in the PDR, together with the increase in far-UV flux ($G_0$) due to keeping U constant, implies stronger PDR [C{\sc ii}] emission with approximately unchanged [N{\sc ii}] emission; which causes an increase in [C{\sc ii}]/[N{\sc ii}] ratio as a function of model density.

Another important effect controlling the line ratio is the conversion of [N{\sc ii}] to [N{\sc iii}] in high ionization conditions (caused by both high U and hard radiation). Its effects can be seen in isolation from the difference between the two ages of the burst of star formation (a proxy for hardness of the incident radiation) since we find that [N{\sc iii}] is very important in the 1 Myr case, while it is negligible in the 10 Myr models.
Its main contribution is clear: while in the low-hardness case an increase in U favors [N{\sc ii}] emission over PDR tracers (like [C{\sc ii}]), this trend reverses for hard radiation, when the  increased [N{\sc iii}] implies a deficit of [N{\sc ii}] emission with increasing ionization parameter.

We have also run equivalent Cloudy simulations with stellar radiation input produced by the lowest available stellar metallicity Geneva tracks to investigate the approximate effect of a young stellar population, as expected at high redshift. The effects on the line ratio of interest do not significantly affect our interpretation ($\lesssim 50\%$). However it is possible that an inclusion of more realistic stellar population models, e.g., accounting for stellar binary effects which could produce higher effective stellar temperatures, may have a greater impact, and requires further investigation. We have not attempted changing the gas-phase metallicity because this would further complicate the interpretation of the model results (e.g., given the uncertainty in the effect of metallicity on the carbon-to-nitrogen ratio).
It is also possible that the structure of the ISM might be significantly altered by radically different stellar populations and star-forming conditions, which would therefore require changing the standard geometry of H{\sc ii} regions around recently formed stars, surrounded by atomic PDR gas, assumed here, which could systematically change the model results. For example, \cite{V16} explore the effect on FIR emission lines of time-dependent cloud photo-evaporation, and suggest that equilibrium models might be systematically incorrect.

\section{Analysis of the individual sources}
\label{analysis}

\subsection{AzTEC-3}

AzTEC-3 is one of the most intensely star-forming starbursts known at $z>5$ (not amplified by gravitational lensing). It is centrally located in one of the earliest galaxy proto-clusters known \citep[showing an 11-fold over-density of LBGs and color-selected galaxies within $2\,$Mpc, suggesting a halo mass $>4 \times 10^{11}\,M_{\odot}$;][]{C11} and is known to be  compact and dusty \citep{R10a,R14}. 
Its very compact starburst ($\lesssim 2.5\,$kpc in diameter), leads to a high dust temperature ($\sim50$--$90\,$K; R14), indicating extreme conditions even among high redshift SMGs.
We have detected the [N{\sc ii}] line in AzTEC-3, with a measured [N{\sc ii}] luminosity of $(2.9\pm1.0) \times 10^8 \,L_\odot$ implying $L_{\rm [NII]}/L_{\rm FIR}=(2.6\pm 1.0) \times 10^{-5}$, which is very low compared to local (U)LIRGs \citep{Z16}.\footnote{The error bars here and in the following express standard percentiles of the appropriate probability distributions.} \cite{L15} and \cite{Z13} argue that low values of [N{\sc ii}] luminosity to FIR emission correlate with warm dust in local galaxy samples. This is consistent with the intense radiation field and high dust temperature already found by SED modeling of AzTEC-3 (R14).

The [N{\sc ii}]/FIR deficit can be  interpreted as the product of two effects which can be studied in isolation: the [C{\sc ii}]/[N{\sc ii}] ratio, which is dependent on the relation between ionized and neutral gas; and the [C{\sc ii}]/FIR ratio, which characterizes the gas heating efficiency. The latter was discussed by R14, where it was interpreted as analogous to the deficit found in local ULIRGs (e.g., \citealt{L98,L03}). On the other hand, the [N{\sc ii}] line in AzTEC-3 also appears marginally fainter compared to [C{\sc ii}] than observed in local (U)LIRGs \citep{F13,K16,Ro15,DS,Z16} and in other high redshift SMGs (\citealt{D14}\footnote{See Section~\ref{discuss} for updated values of  [C{\sc ii}]/[N{\sc ii}] in the objects studied by \cite{D14}.}; \citealt{B1}), for which typical sample averages of $L_{\rm [CII]}/L_{\rm [NII]}$ are in the range $\sim10-20$.
In addition, the measured $L_{\rm [CII]}/L_{\rm [NII]}$ ratio of $22\pm8$ might be underestimated  due to optical depth effects.  R14 found that dust extinction might significantly affect the [C{\sc ii}] flux (we estimate a factor of $\sim1.6$ in the line ratio, based on the global SED-modeling parameters from R14) and could obstruct the view to the innermost, densest regions.

On the other hand, if we qualitatively extrapolate the correlation between star-formation rate surface density and ionized gas density and ionization parameter found by \cite{HC16} in local galaxies, we are led to expect gas densities $\sim 1000\,$cm$^{-3}$ in the H{\sc ii} regions in the central part of AzTEC-3, far above the [N{\sc ii}] $205\,\mu$m critical density of $100\,$cm$^{-3}$. Based on our Cloudy simulations such high gas densities in the H{\sc ii} regions would produce a [C{\sc ii}]/[N{\sc ii}] ratio almost an order of magnitude higher than observed (Fig.~\ref{Cloudy}). The finding of a ``normal" ratio of 20--40 is therefore surprising and needs to be interpreted carefully.

\cite{R16} found a [C{\sc ii}]/[N{\sc ii}] luminosity ratio of 12 in the central regions of IC 342, a normal local star-forming galaxy, similar to our own Milky Way, suggesting that a majority of the [C{\sc ii}] emission originates in ionized gas. However, \cite{L15} find a positive correlation between the [C{\sc ii}]/[N{\sc ii}] ratio and bluer FIR-color in local galaxies. Based on the expectation that FIR-color correlates both with ionization parameter and with ionized gas density \citep{HC16}, we interpret this trend as predominantly a density effect, reducing the importance of the ionized gas component of [C{\sc ii}] emission in hotter, denser starbursts. 

Why does the trend to higher [C{\sc ii}]/[N{\sc ii}] ratios not extend to high redshift SMGs, and especially AzTEC-3 with its extreme density compared to local ULIRGs?
It is possible that the tentatively detected low-level, extended emission, if real, might indicate a warm ionized component that is more extended and diffuse than the denser star-forming, FIR and [C{\sc ii}]-emitting central region. This may also indicate a gas flow either in the process of being ejected or accreting onto the central galaxy.
Furthermore, the  tentative evidence for a broader [N{\sc ii}] line than [C{\sc ii}], (660 and $410\,$km s$^{-1}$ respectively) could be indicative of higher dynamical mass enclosed within the [N{\sc ii}]-tracing gas versus the neutral star-forming medium or it could be associated with the systemic velocity in a potential ionized gas flow.

This tentative evidence for non-nuclear emission may suggest that, in contrast to local starbursts, the hard and intense radiation field of AzTEC-3 might ionize a dominant large diffuse component of the ISM, outside the dense, molecular star-forming core. The lower density of this component, perhaps comparable to the critical density of [N{\sc ii}], would imply that this  gas would dominate the emission over the gas directly surrounding the current star-formation. Higher signal-to-noise observations are necessary to further investigate this possibility.

The slightly elevated [C{\sc ii}]/[N{\sc ii}] ratio in AzTEC-3, compared to other high-$z$ SMGs, is unlikely to be due to gas-phase metallicity alone since this source is very dusty ($M_{\rm d}=2.7\times 10^8\,M_\odot$; R14). The low dynamical mass also puts an upper limit on $\alpha_{\rm CO}<1.3\;{\rm M_{\odot}(K\,km\,s^{-1}\,pc^2)^{-1}}$ (R14), which is incompatible with expectations for low-metallicity gas \citep{B2}.

If the [N{\sc ii}] emission is indeed extended at very faint levels, perhaps over a scale up to $\sim15\,$kpc, our results may be suggestive of ionized gas flows, either ejected from the hyper-starburst activity or funneling gas into the core of the dense proto-cluster associated with AzTEC-3, fueling the extreme star-formation.
The strong OH molecular emission feature detected in R14, apparently blue-shifted, is indicative of warm, dense ouflows which could be related to the tentatively detected extended ionized gas, if confirmed by more sensitive observations.

Assuming that the [N{\sc ii}]-emitting ionized gas is also emitting [C{\sc ii}] with a predictable relative intensity, we can determine the fraction of [C{\sc ii}] emission that comes from ionized rather than neutral gas using the measured [C{\sc ii}]/[N{\sc ii}] ratio.\footnote{This particular inference depends on several assumptions. The line emission needs to be optically thin in order to be additive. Also, the ratio of carbon and nitrogen gas-phase abundances are not well constrained and enter the conversion of [N{\sc ii}] to [C{\sc ii}] luminosity coming from ionized gas. In addition, complete and equivalent ionization state of carbon and nitrogen in the traced ionized gas is assumed (i.e., we assume that in N$^{++}$ regions carbon atoms are also in a higher ionization state, thus not contributing to the [C{\sc ii}] emission). Furthermore, the inferred fraction of [C{\sc ii}] emission from ionized gas represents a spatial average over possibly heterogeneous regions.} Following \cite{O06}, we assume a C$^+$/N$^+$ abundance ratio of $\sim1.8$  \citep[from][]{S96}, to infer a line luminosity ratio [C{\sc ii}]$_{\rm ionized}$/[N{\sc ii}]$\simeq$3.5, for ionized gas density in the range $\sim$10--$1000\,$cm$^{-3}$. With these assumptions the measured fraction of [C{\sc ii}] emission from ionized gas in AzTEC-3 lies in the range 10--25\%. 

Following \cite{Fa16b} we can use the {\it Spitzer} IRAC [$3.6\,\mu$m]--[$4.5\,\mu$m] color to qualitatively constrain the strengths of some of the main optical emission lines. At the redshift of AzTEC-3 the H$\alpha$ line falls into the $4.5\,\mu$m band, and therefore the {\it Spitzer} [$3.6\,\mu$m]--[$4.5\,\mu$m] color is related to its H$\alpha$ equivalent width. From the COSMOS2015 catalog \citep{Lcosm16} we calculate a color of $0.48\,$mag, close to the average for the $z=5$--6 galaxy population \citep{Fa16b}. However the optical counterpart of AzTEC-3 is shifted with respect to the (sub)millimeter continuum emission, indicative of strong dust obscuration \citep[e.g.,][R14]{R10a}; hence it is possible that the optical emission lines may not be directly associated with the peak starburst emission, but either with a secondary component or with a less obscured part of the central galaxy.






\subsection{HZ10}

We have detected [N{\sc ii}] in the most IR-luminous LBG of the C15 sample, HZ10 ($z\sim5.7$). This galaxy appears to be more dusty than average LBGs at its redshift based on our ALMA continuum measurements at $\sim158\,\mu$m and $\sim205\,\mu$m. It displays extended [C{\sc ii}] emission, showing a smooth velocity gradient possibly suggestive of coherent rotation and is characterized by a $L_{\rm [CII]}$/$L_{\rm FIR}$ ratio of $1.5 \times 10^{-3}$ (C15). This ratio is suggestive of ``normal" star-forming conditions (radiation fields and density perhaps similar to local star-forming galaxies), which is surprising for a young stellar population expected at $z=5$--6.
The measured [N{\sc ii}] luminosity of $(2.1\pm0.8) \times 10^8 \, L_\odot$ corresponds to $L_{\rm [NII]}/L_{\rm FIR}=(3.0^{+4.7}_{-1.9}) \times 10^{-4}$, when using our median estimate for the FIR luminosity, placing it in the regime of local star-forming galaxies \citep{F13,K16,Ro15,DS,Z16}. 

In contrast to AzTEC-3, the extended gas reservoir in HZ10 suggests a lower density star-forming environment ($\Sigma_{\rm SFR}\sim 1/50\times$ in AzTEC-3, based on a comparison of continuum sizes and SFRs).  This lower $\Sigma_{\rm SFR}$ is qualitatively consistent with the lower [C{\sc ii}]/[N{\sc ii}] ratio observed in HZ10, but it appears not to cause as large an effect as we might expect from our Cloudy models, which is consistent with the interpretation that a significant fraction of the [N{\sc ii}] emission in AzTEC-3 is probably dominantly emitted by a more diffuse medium than the actively star-bursting gas.
The global [C{\sc ii}]/[N{\sc ii}] ratio of $\sim20$ measured for HZ10 is comparable to local  galaxies with high SFRs \citep{Z16}, indicating a significant contribution of ionized gas to the [C{\sc ii}] emission. In fact, under the same assumptions as for AzTEC-3, we find a fraction of [C{\sc ii}] emission from ionized gas in HZ10 in the range 10--25\%, which is similar to what is found in ``normal" local star-forming galaxies. This, together with our result for LBG-1, would support the interpretation that the ISM in HZ10 is more mature than in ``normal" LBGs at $z=5$--6.  Further evidence in favor of this interpretation comes from the stellar mass of HZ10 ($\sim2\times 10^{10}\,M_\odot$; C15) being higher than in most of the C15 sample, suggesting that HZ10 may have synthesized more metals already, consistent with the finding of a more mature ISM.
Furthermore, the ionized fraction could be enhanced in parts of the galaxy; the tentative local line ratios determined with our [N{\sc ii}]-based ``red" and ``map peak" methods (see Section~\ref{resHZ10} and Fig.~\ref{spectra}) suggest a [C{\sc ii}]/[N{\sc ii}] ratio as low as $\sim10$, perhaps caused by an ionization-dominated region of the galaxy. Such regions are consistent with what is seen in the models of \cite{V13}. The tentative evidence for heterogeneity of the ionization state in the star-forming gas reservoir of HZ10 is suggestive of regions of different density (clumpiness), a recent massive starburst event, or a recent/active merger.

At the redshift of HZ10, the H$\alpha$ line falls into the $4.5\,\mu$m band, and the [O{\sc iii}] and H$\beta$ lines fall into the $3.6\,\mu$m band, allowing us to use the relative equivalent widths of the two sets of lines as a diagnostic tool \citep{Fa16b}; the [$3.6\,\mu$m]--[$4.5\,\mu$m] color in this case is a measure of the strength of the [O{\sc iii}] emission line relative to the hydrogen lines. In HZ10, we measure a [$3.6\,\mu$m]--[$4.5\,\mu$m] color of $0.41\,$mag, which falls towards the ``red" end of the distribution for comparable galaxies, implying that [O{\sc iii}] is not as bright as observed in normal galaxies at the same redshift. This lower than average [O{\sc iii}] strength is consistent with our inference of a more mature galaxy, which is less dominated by high ionization state gas than LBG-1.

\begin{figure*}[tbh]
\hspace{-15pt}\includegraphics[scale=0.5]{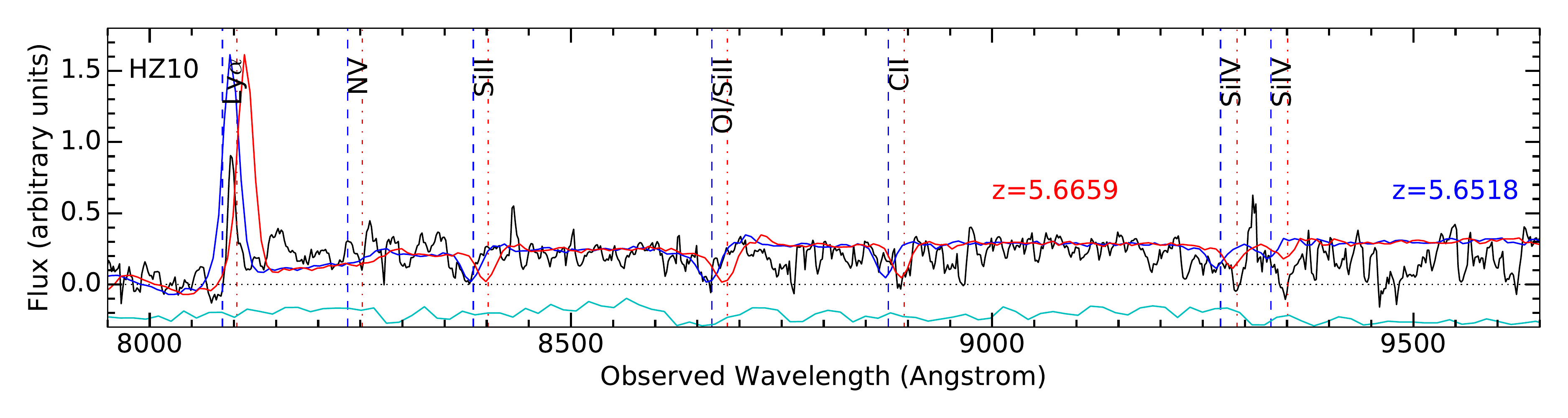}
\caption{Optical Keck/DEIMOS spectrum of HZ10 shown in black with UV emission/absorption lines identified for two different redshifts ($z=5.6518$ and 5.6659). Si {\sc ii}, O {\sc i}/Si {\sc ii} and Ly$\alpha$ are slightly blue-shifted ($110\pm180\,$km s$^{-1}$) relative to the systemic redshift inferred from the FIR [C{\sc ii}] line, likely due to outflowing stellar winds. The UV C {\sc ii} absorption and the high-ionization Si {\sc iv} lines are redshifted with respect to the systemic velocity ($522\pm200\,$km s$^{-1}$) and are compatible with the FIR [N{\sc ii}] emission redshift, which may indicate higher ionization conditions in a part of the galaxy. The blue (red) spectrum shows the stacked $z\sim3$ LBG template from \cite{Sh03} redshifted to the best fitting match to the two observed sets of lines in HZ10 and scaled to its continuum flux level. The cyan line shows the telluric transmission spectrum (arbitrarily scaled), estimated by the inverse sample variance.}
\label{spec_opt}
\end{figure*}

Fig.~\ref{spec_opt} shows an optical (rest-UV) spectrum of HZ10, taken with Keck/DEIMOS as part of the COSMOS DEIMOS spectroscopic survey (PI: Capak). This spectrum shows Ly$\alpha$ emission and some of the main UV absorption lines commonly observed in LBGs \citep{Sh03}; we identify Si {\sc ii}, O {\sc i}/Si {\sc ii}, C {\sc ii}, N {\sc v} and Si {\sc iv}. Assuming that the FIR [C{\sc ii}] emission traces the systemic velocity for the whole galaxy, several low-ionization ISM lines (Si {\sc ii}, O {\sc i}/Si {\sc ii} and Ly$\alpha$) are slightly blue-shifted by $(110\pm180)\,$km s$^{-1}$ relative to the systemic velocity, likely due to outflowing stellar winds, as commonly observed in $z\sim 3$ LBGs \citep{Sh03}. Interestingly, the C {\sc ii} absorption (usually originating from the neutral ISM) and the high-ionization Si {\sc iv} lines (mostly originating from H{\sc ii} regions and stellar winds) are significantly redshifted with respect to the systemic velocity by $(522\pm200)\,$km s$^{-1}$. This redshift, consistent with what is observed for the FIR [N{\sc ii}] line ($255\pm50\,$km s$^{-1}$), may suggest a potential common origin for these lines, which thus may predominantly emerge from the western part of the galaxy.  This observation would be consistent with the velocity gradient seen in the FIR [C{\sc ii}] emission, which shows a spatial displacement from west to east between the red- and blueshifted part of the emission, indicating a higher ionization fraction in the western part of HZ10 compared to the rest of the galaxy.

\subsection{LBG-1}

We tentatively detect [N{\sc ii}] emission from LBG-1, a relatively quiescent star-forming galaxy in the AzTEC-3 proto-cluster that can be considered ``typical" of LBGs at $z=5$--6  based on its UV properties (luminosity, spectral slope) and low IR luminosity (R14). LBG-1 shows a complex, three-component morphology with velocity gradients in the ALMA [C{\sc ii}] data confirming the physical association of the HST sources, and suggesting either a merger or an extended clumpy star-forming gas reservoir.
LBG-1 shows an extremely faint FIR continuum compared to its [C{\sc ii}] emission ($L_{\rm [CII]}/L_{\rm FIR}\sim 10^{-2}$), pointing towards low dust content and perhaps low gas-metallicity conditions that are suggestive of a young starburst, possibly triggered by the merging/accreting event responsible for the complex morphology.
The potential absence of an older stellar population makes this galaxy an ideal target for exploring the state of the ISM in young, forming galaxies typical in the first giga-year of cosmic time.

The measured [N{\sc ii}] luminosity of $(2.5\pm1.9) \times 10^7 \,L_\odot$ corresponds to a ratio of $L_{\rm [NII]}/L_{\rm FIR}=(1.1^{+4.0}_{-1.0}) \times 10^{-4}$, which is lower than what we find in HZ10, suggestive of  different ISM properties.

In particular, the  [C{\sc ii}]/[N{\sc ii}] ratio is very high ($68^{+200}_{-28}$) compared to local normal star-forming galaxies as well as all high redshift galaxies observed to date, indicating that only a small fraction of the bright [C{\sc ii}] emission is coming from ionized gas. 
It is unlikely that the mechanism responsible for the high ratio of [C{\sc ii}]/[N{\sc ii}] is the same as in AzTEC-3 since LBG-1 has a low star-formation rate and spatially extended [C{\sc ii}] emission.
At least two scenarios are compatible with our observations: either the ISM is dominated by neutral PDR gas, or most of the ionized gas could be in a high ionization state (compared to Galactic H{\sc ii} regions) with a larger abundance of $\rm N^{++}$  than $\rm N^+$.

A low dust-to-gas ratio could imply very extended PDRs, where the transition to molecular hydrogen is due to self-shielding instead of dust opacity. This interpretation is consistent with a large neutral-to-ionized gas fraction by increasing the proportion of the atomic regions emitting in [C{\sc ii}].

As suggested by $z\sim2$--3 studies \citep[e.g.,][]{K13} we assume a large ionization parameter to be characteristic of normal star-forming galaxies, especially at this higher redshift. The trend to harder radiation fields and low dust content could cause the ISM properties to differ dramatically in this extremely young star-forming galaxy, which suggests that we may have to turn to local dwarfs in the search for local analogs.
In the case of dwarf galaxies, \cite{Co15} also find extremely elevated values for the [C{\sc ii}]/[N{\sc ii}] ratio, although this is deduced indirectly as they only measured [N{\sc ii}] $122\,\mu$m, which has a higher critical density.
They also find evidence for strong [O{\sc iii}] emission, which is indicative of higher ionization in the majority of the ionized gas; they associate the high ionization with the deficit in [N{\sc ii}] emission. \cite{DL14} and \cite{Co15} also find the [O{\sc iii}] emission to be the dominant FIR line by luminosity, while local H{\sc ii} regions are almost never dominated by such high-ionization gas.
This explanation could be applicable to LBG-1 and would support the need to invoke a low dust content, as in dwarfs, to explain the low [N{\sc ii}] luminosity.

At the present level, a third interpretation such as a lower nitrogen abundance in the gas phase remains possible. A reduced gas phase metallicity would not directly affect the [C{\sc ii}]/[N{\sc ii}] ratio through the ion abundances unless it were differentially reducing nitrogen relative to carbon. Local studies \citep[e.g.,][]{Vi16} have shown a non-linear dependence of the nitrogen abundance relative to oxygen and perhaps carbon, and low metallicity gas could be expected to show reduced nitrogen abundance if it behaves as a secondary element. On the other hand, no conclusive evidence has been found so far at high redshift, with results indicating a potentially higher nitrogen abundance than expected or a N/O ratio that depends on stellar mass in tracing galaxy maturity \citep[e.g.,][]{M16,S14}.

Under the same assumptions as for AzTEC-3, we find a fraction of [C{\sc ii}] emission from ionized gas in LBG-1 that is only about $\sim5\%$. We do not expect the presence of significant [N{\sc iii}]-emitting regions to affect this conclusion as the higher ionization-state gas is not expected to emit [C{\sc ii}].

We measure a {\it Spitzer} IRAC [$3.6\,\mu$m]--[$4.5\,\mu$m] color of $0.44\,$mag for LBG-1 which is consistent with a median value of the H$\alpha$ equivalent width for LBGs, i.e., consistent with what is expected for a ``typical" galaxy at this redshift \citep{Fa16b}. The [O{\sc iii}]/H$\beta$ lines are not redshifted into the $3.6\,\mu$m window for LBG-1.

\begin{figure}[thb]
\hspace{-20pt}\includegraphics[scale=.4]{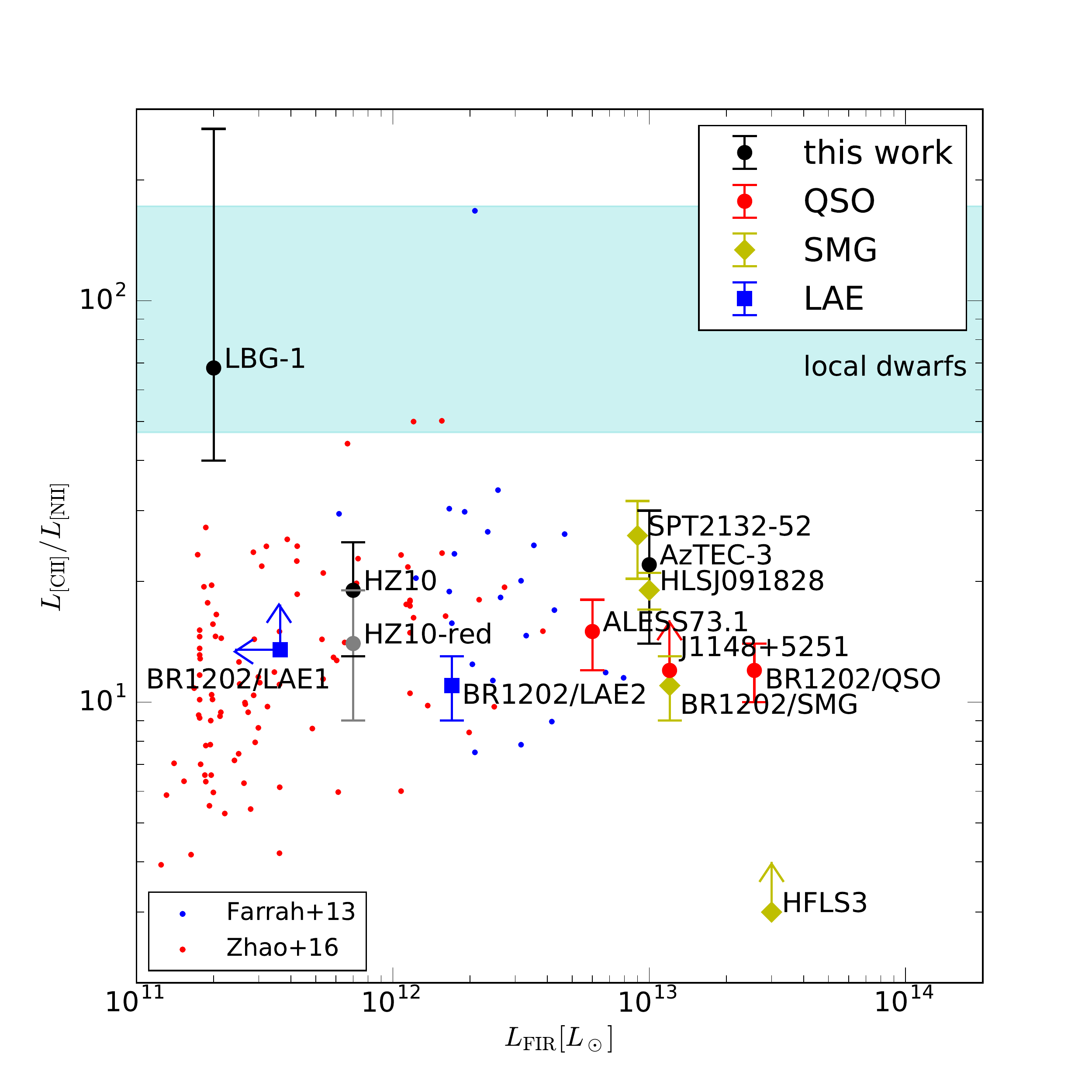}
\caption{[C{\sc ii}]/[N{\sc ii}] line luminosity ratios observed in high redshift galaxies to date as a function of their FIR luminosity (\citealt{CW13,dB14,B1, Rw14, Co12, C13,R13}; R14; C15). For comparison we also show a sample of ULIRGs from \cite{F13} (using the [N{\sc ii}] $122\,\mu$m line) and LIRGs with [N{\sc ii}] (using the [N{\sc ii}] $205\,\mu$m line) from \cite{Z16} and [C{\sc ii}] from \cite{DS}. The range of ratios in dwarfs \citep{Co15} (using the [N{\sc ii}] $122\,\mu$m line) is shown as a cyan band. The [N{\sc ii}] $122\,\mu$m line measured in the indicated local samples was converted to a [N{\sc ii}] $205\,\mu$m luminosity assuming a ratio of 1/2.5, estimated from \cite{HC16}. The abscissa in the local samples are defined as total IR luminosity; no attempt was made to convert to a common FIR luminosity scale because it does not affect our interpretation.}
\label{ratio_plot}
\end{figure}

\section{Discussion and Conclusions}
\label{discuss}

We have detected  [N{\sc ii}] $205\,\mu$m emission towards the highest redshift sample of galaxies to date. Our sample spans almost two orders of magnitude in star formation rate and includes the compact starbursting SMG AzTEC-3 ($z=5.3$; SFR$\,{\rm\sim1100\,M_{\odot}\,yr^{-1}}$; R14). From the combined  R14 \& C15 sample of [C{\sc ii}]-detected LBGs we also selected the relatively dusty, above average SFR galaxy, HZ10 ($z=5.7$; SFR$\,{\rm\sim170\,M_{\odot}\,yr^{-1}}$), which lies at the upper limit, although within the scatter of the $z=5$--6 star-forming Main Sequence, and a typical lower-SFR galaxy, LBG-1 (SFR$\,{\rm\sim10-30\,M_{\odot}\,yr^{-1}}$),  which we consider representative of the ``normal" population of star-forming galaxies at $z\sim5.3$ because it appears to lie on the Main Sequence. 

Our observations of the [C{\sc ii}]/[N{\sc ii}] luminosity ratio are summarized in Fig.~\ref{ratio_plot}, and are shown together with all other high-redshift measurements to date, as well as some local galaxy samples.
We have re-measured the [N{\sc ii}] line luminosity in the BR1202--0725 system members (a QSO, an SMG and two LAEs at $z=4.7$) based on  higher sensitivity archival ALMA data (project ID: 2013.1.00745.S) to update the ratios presented in \cite{D14}. The [N{\sc ii}] line is detected with high significance for the QSO, SMG, and LAE-2, but is not detected in LAE-1.\footnote{The [N{\sc ii}] line fluxes are $1.5 \pm 0.2 \,$Jy km s$^{-1}$ for the SMG,  $0.74 \pm 0.07\,$Jy km s$^{-1}$ for the QSO,  $<0.5\,$mJy peak ($3\sigma$ limit) for LAE-1, and  $0.30 \pm 0.06\,$Jy km s$^{-1}$ for LAE-2.} The non-detection of [N{\sc ii}] emission in LAE-1 may not be strongly constraining if it is similar to LBG-1, with the caveat that it is likely to be affected by tidal forces from the SMG-QSO merger and the intense QSO radiation.  Furthermore, the [N{\sc ii}] line in LAE-2 is very broad ($\sim1000\,$km s$^{-1}$), suggesting that the [C{\sc ii}] line measured in \cite{C13} is truncated due to the band edge, and thus may only cover $\sim$30\% of the full line emission.\footnote{Adopting the [C{\sc ii}] line luminosities from \cite{C13}, the [C{\sc ii}]/[N{\sc ii}] line ratios are  $11\pm2$ for the SMG, $12\pm2$ for the QSO, $>13.5$ for LAE-1, and, when using the velocity range corresponding to the [C{\sc ii}] measurement, $11\pm2$ for LAE-2.} The very broad [N{\sc ii}] line in LAE-2 is compatible with the Ly$\alpha$ FWHM of $1225\pm257\,$km s$^{-1}$ in \cite{W14}, which may imply that the ISM in this galaxy is strongly affected by quasar winds.

Recent work \citep[e.g.,][]{K13,S16} suggests that stellar metallicity in early galaxies might be very low compared to gas-phase metallicity (which is dominated by oxygen, nitrogen, and carbon abundances). This is due to the expected enrichment time for iron of the order of $1\,$Gyr; the iron abundance is responsible for the biggest contribution to stellar opacity and is not enriched in the ISM by core-collapse supernovae and is therefore expected to be low in galaxies at this epoch, hence greatly affecting massive star structure and UV radiative output. Since we expect the gas properties of the ISM under investigation to depend on the hardness and intensity of the stellar radiation field, these properties may not be directly comparable to local galaxies. It is also expected that the high gas fractions typical for galaxies at these epochs might strongly modify the gas phase structure \citep[e.g.,][]{V13}.

Figure~\ref{ratio_plot} shows that there seems to be a relatively narrow range (0.5 dex) for  [C{\sc ii}]/[N{\sc ii}] ratios in SMGs and quasar-hosts at high-redshift. Their range is similar to, although perhaps slightly higher than, the ratios measured in local LIRGs and ULIRGs. 
This does not fit our expectations based on local trends with density and star formation rate surface density \citep[e.g.,][]{L15}; these trends would suggest a significantly higher ratio than we observe.
We therefore suggest that the global measurements at high-redshift may not be observing the [N{\sc ii}] emission coming from the equivalent regions in local starbursts, where the H{\sc ii} regions at the site of recent star-formation dominate, but a more diffuse ISM component which might not be as prevalent in less gas-rich local analogs. If this were to be correct, it is possible that the ionized gas in the nuclear regions, all of which is surrounding the recently formed massive stars may be in a state of higher ionization.  In this case both nitrogen and oxygen may predominantly be in their doubly ionized state as perhaps observed in LBG-1.

It is likely that in order to unravel the physical origin of the line ratio under consideration, spatially resolved line studies are needed; spatial global averages might give a very biased view of the physical gas conditions if the neutral and the ionized gas are not co-spatial.  HZ10 is the first high redshift galaxy where this type of study seems possible. In fact, we find tentative evidence for the [N{\sc ii}] emission originating from only part of the [C{\sc ii}]-emitting region, which is reminiscent of clumpy star-forming gas disks observed in other tracers of ionized gas like H$\alpha$ \citep[e.g.,][]{G11,W15}.
Since rest-frame optical wavelength tracers like H$\alpha$ are presently not accessible in the first giga-year of cosmic time, FIR fine-structure lines like [N{\sc ii}] may become the tracer of choice for this kind of resolved clump/disk dynamic studies in the earliest galaxies.
Due to the lower density and dust content relative to a compact starburst like AzTEC-3, regions of different ionization might be more easily accessible in less extreme $z>5$ galaxies, like HZ10. Future studies are necessary to determine if the regions with lower [N{\sc ii}] emission are dominated by neutral gas, perhaps at higher density than the [N{\sc ii}]-emitting fraction (as potentially observed in AzTEC-3), or whether [N{\sc iii}] dominates the nitrogen emission in these regions, suggesting LBG-1-like conditions.

A different effect may become dominant for ``normal"  galaxies at $z=5$--6, like LBG-1, where the low dust content of the ISM and the very young stellar population perhaps start affecting the phase structure of the low-metallicity gas. In fact, our data constrain the [C{\sc ii}]/[N{\sc ii}] ratio to be larger than $\sim 40$ and probably in the 60--100 range, indicative of a minor contribution from ionized gas to [C{\sc ii}] compared to that in local normal star-forming galaxies, for which the measured ratio is typically 10--25 \citep{Z16,DS}. Understanding the [C{\sc ii}] emission line in such galaxies appears crucial as the low FIR and CO emissions (R14, C15) suggest that [C{\sc ii}]  might be an essential neutral gas coolant  in low dust environments at $z>5$, which is necessary to achieve dense gas formation and hence star-formation. Our measurement is compatible with the recent findings of high ratios in local dwarfs \citep{Co15}, strengthening the case that these might be appropriate analogs for the ISM in early galaxies, in some respects.
\cite{Co15} find [O{\sc iii}] to be the brightest FIR line in these galaxies, suggesting abundant ionized gas in a higher ionization state than what is traced by [N{\sc ii}]. Measuring either  [O{\sc iii}]  or  [N{\sc iii}] will be necessary to determine the relative importance of the ionized ISM component in normal $z=5$--6 galaxies, and it could become an important tool due to its brightness, possibly comparable to [C{\sc ii}] or brighter \citep{I16}. The extremely low [N{\sc ii}] value measured in LBG-1 holds further evidence in favor of hard and intense radiation in early star-forming galaxies, with possibly much higher escape fraction of hard ionizing photons (dominating the ISM throughout the galaxy) due to low dust abundances, and therefore could potentially help to shed light on the mechanisms of cosmic re-ionization.

\smallskip
We thank the anonymous referee for a helpful and constructive report and T. K. Daisy Leung for helpful discussion. DR and RP acknowledge support from the National Science Foundation under grant number AST-1614213 to Cornell University. R.P. acknowledges support through award SOSPA3-008 from the NRAO. AK acknowledges support by the Collaborative Research Council 956, sub-project A1, funded by the Deutsche Forschungsgemeinschaft (DFG).
DR acknowledges the hospitality at the Aspen Center for Physics and the Kavli Institute for Theoretical Physics during part of the writing of this manuscript.
The National Radio Astronomy Observatory is a facility of the National Science Foundation operated under cooperative agreement by Associated Universities, Inc.
This paper makes use of the following ALMA data: ADS/JAO.ALMA\#2015.1.00928.S, 2011.0.00064.S, 2012.1.00523.S, 2011.0.00006.SV and 2013.1.00745.S. ALMA is a partnership of ESO (representing its member states), NSF (USA) and NINS (Japan), together with NRC (Canada), NSC and ASIAA (Taiwan), and KASI (Republic of Korea), in cooperation with the Republic of Chile. The Joint ALMA Observatory is operated by ESO, AUI/NRAO and NAOJ




\end{document}